\documentstyle[12pt]{article}
\newcommand{\ket}[1]{\left | #1 \right \rangle}

\begin{document}
\noindent
{\small Appearing in {\em Phil. Trans. Roy. Soc. (Lond.)} 1998,
Proceedings of Royal Society Discussion Meeting ``Quantum
Computation: Theory and Experiment'' held in November 1997.}

\begin{center}
{\large\bf Quantum Algorithms: Entanglement Enhanced \\Information
Processing}
\bigskip
\\ Artur Ekert\\Clarendon Laboratory\\ University of Oxford\\
Parks Road, Oxford OX1 3PU, England.\\ \bigskip
 Richard Jozsa\\
School of Mathematics and Statistics\\ University of Plymouth\\
Plymouth, Devon PL4 8AA, England.

\end{center}
\bigskip
{\bf Abstract:} {\em We discuss the fundamental role of
entanglement as the essential nonclassical feature providing the
computational speedup in the known quantum algorithms. We review
the construction of the Fourier transform on an Abelian group and
the principles underlying the fast Fourier transform algorithm. We
describe the implementation of the FFT algorithm for the group of
integers modulo $2^n$ in the quantum context, showing how the
group-theoretic formalism leads to the standard quantum network and
identifying the property of entanglement that gives rise to the
exponential speedup (compared to the classical FFT). Finally we
outline the use of the Fourier transform in extracting
periodicities, which underlies its utility in the known quantum
algorithms.
 }\\[5mm] {\large\bf
Introduction}\\[3mm] In 1982 Feynman\cite{FE} noted a profound
difference in the nature of physical evolution governed by the laws
of quantum physics as compared to evolution under the laws of
classical physics. He observed that quantum mechanics (apparently)
cannot be {\em efficiently} simulated on a classical computer (or
by any classical means) i.e. that the simulation of a general
quantum evolution by any classical means appears to involve an
unavoidable exponential slowdown in running time. This observation
embodies the essence of the subject of quantum computation so we
will begin by elaborating its meaning in terms of a simple example.

Consider a discrete sequential quantum process defined as follows.
We start with a row of qubits (i.e. 2 level systems with a
preferred basis labelled $\{ \ket{0}, \ket{1} \}$) all initially in
state $\ket{0}$:
\[ \begin{array}{cccccc}
\ket{0} & \ket{0} & \ket{0} & \cdots & \ket{0} & \cdots \nonumber \\
\mbox{qubit 1} & \mbox{qubit 2} & \mbox{qubit 3}  & \cdots & \mbox{qubit j} & \cdots
\nonumber \end{array} \]
We are also given a fixed 2-qubit interaction (or 2-qubit ``quantum
gate'') $U$ which is a unitary operation that may be applied to any
selected pair $(i,j)$ of qubits. Furthermore we have a program of
instructions specifying the pairs of qubits to which the gate
should be sequentially applied. Thus step $k$ of the program is
``apply $U$ to qubits $(i_k ,j_k )$ and replace them in the row'',
for $k=1,\ldots ,n$.    After $n$ steps in this process we measure
qubit 1 in its preferred basis obtaining 0 or 1 according to a
probability distribution ${\cal P}_n = \{ p_n (0), p_n (1) \}$.
Thus by implementing this process in an actual quantum physical
system we can sample the distribution ${\cal P}_n$ in time $O(n)$
i.e. after a time which grows linearly with $n$.

Our problem is to mimick this process by classical means. More
precisely, we wish to describe a classical probabilistic process
which enables us to sample the distribution ${\cal P}_n$ defined by
the above quantum process. A simple way of achieving this is the
following. The operation $U$ is just a $4\times 4$ unitary matrix
and given the starting state with the program, we can sequentially
compute by hand -- using simple matrix multiplication -- the
quantum state at each successive stage $k$. Then knowing the state
at stage $n$, the rules of quantum measurement theory enable us to
calculate $p_n (0)$ and $p_n (1)$, so finally we toss a
correspondingly biassed coin.

This classical simulation has the following notable characteristic
feature. The quantum state after $k$ steps is generally a $k$-qubit
entangled state requiring $O( 2^k )$ coefficients for its
description (as typically an extra qubit may be brought in at each
step). Thus because of entanglement, we have an exponential growth
in time of the information needed to describe the state. Hence the
classical simulation will slow down exponentially in time under the
weight of this exponentially growing information that needs to be
processed in each step. To sample ${\cal P}_n$ our classical
simulation will require $O(2^n )$ time  while the quantum process
marches ahead in unflagging {\em linear} time. There is no more
efficient classical method known to solve this problem. Thus
according to the laws of quantum mechanics, Nature remarkably is
able to process information exponentially more efficiently than can
be achieved by any classical means!

Note that if only {\em product} states of qubits were available
then the information needed to describe the state would grow only
{\em linearly} with $n$ (being $n$ times the amount of information
needed to describe a typical single qubit state). Thus the
exponential speedup in our example of quantum information
processing is fundamentally a feature of quantum {\em
entanglement}. This point has been elaborated in Jozsa\cite{J1}.
Indeed it provides an extraordinary manifestation of entanglement
which is entirely independent of the auxiliary notion of
non-locality.

We may attempt to mimick the quantum process using classical {\em
waves} which admit the possibility of {\em superposition} of modes.
For example we might represent each qubit by a vibrating elastic
string with fixed endpoints and select two lowest energy modes of
vibration to represent the states $\ket{0}$ and $\ket{1}$. It is
then possible to construct the general superposition corresponding
to $a\ket{0}+b\ket{1}$. However, regardless of how much the strings
interact with each other in their subsequent (externally driven)
vibrational evolution, their joint state is always a {\em product}
state of $n$ separate vibrations. The total state space of the
total classical system is the {\em Cartesian} product of the
individual state spaces of the subsystems whereas
quantum-mechanically, it is the {\em tensor} product. This crucial
distinction between Cartesian and tensor products is precisely the
phenomenon of quantum entanglement. Nevertheless, we may yet
attempt to represent entanglement using classical waves in the
following manner. The state of $n$ qubits is a $2^n$ dimensional
space and can be isomorphically viewed as the state space of a {\em
single} particle with $2^n$ levels. Thus we simply interpret
certain states of a single $2^n$ level particle as ``entangled''
via their correspondence under a chosen isomorphism between
$\bigotimes^n {\cal H}_2 $ and ${\cal H}_{2^n}$ (where ${\cal H}_k$
denotes a Hilbert space of dimension $k$.) In this way, $2^n$ modes
of a classical vibrating system can apparently be used mimick
general entanglements of $n$ qubits. However the physical
implementation of this correspondence appears always to involve an
exponential overhead in some physical resource so that the
isomorphism is {\em not} a valid correspondence for considerations
of complexity i.e. when the amount of physical resources required
to achieve the representation is taken into account. For example
suppose that the $2^n$ levels of the one-particle quantum system or
corresponding classical system, are equally spaced energy levels. A
general state of $n$ qubits requires an amount of energy that grows
{\em linearly} with $n$ (since we will need at most to excite each
qubit to its upper level) whereas a general state of the $2^n$
level quantum or classical system requires an amount of energy that
grows {\em exponentially} with $n$. To physically realise a system
in a general superposition of $2^n$ modes we need exponential
resources classically and linear resources quantum mechanically
{\em because of the existence of entanglement}.

Our discussion above about the information needed to describe a
state, indicates that $n$ qubits have an exponentially larger
capacity to represent information than $n$ classical bits. Note
that although $n$ classical bits have $2^n$ possible states, each
of these states may be described by just $n$ bits, in contrast to
the quantum situation where $O(2^n )$ superposition components may
be involved in a single state. However the information embodied in
the quantum state has a further remarkable feature -- most of it is
{\em inaccessible} to being read by any possible means! Indeed
quantum measurement theory places severe restrictions on the amount
of information that we can obtain about the identity of a given
unknown quantum state. This intrinsic inaccessibility of the
information may be quantified \cite{HOL,FP} in terms of Shannon's
information theory\cite{CT}. In the case of a general state of $n$
qubits, with its $O(2^n )$ information content, it turns out that
at most $n$ classical bits of information about its identity may be
extracted from a single copy of the state by any physical means
whatsoever. This coincides with the maximum information capacity of
$n$ classical bits.

The full (largely inaccessible) information content of a given
unknown quantum state is called quantum information. Natural
quantum physical evolution may be thought of as the processing of
quantum information. Thus the viewpoint of computational complexity
reveals a new bizarre distinction between classical and quantum
physics: to perform natural quantum physical evolution, Nature must
process vast amounts of information at a rate that cannot be
matched by any classical means, yet at the same time, most of this
processed information is kept hidden from us! However it is
important to point out that the inherent inaccessibility of quantum
information does {\em not} cancel out the possibility of exploiting
this massive information processing capability for useful
computational purposes. Indeed, small amounts of information may be
extracted about the overall identity of the final state which would
still require an exponential effort to obtain by classical means.
The ability to sample the probability distribution ${\cal P}_n$
above provides an example. A more computationally useful example is
given by the technique of ``computation by quantum parallelism''
\cite{DD1,DJ} according to which a superposition $\sum_{i=1}^{2^n}
\ket{i}$ of exponentially many input values $i$ for a function $f$
may be set up in linear time and a {\em single} subsequent function
evaluation will provide exponentially many function values in
superposition as $\ket{f}=
\sum_i \ket{i}\ket{f(i)}$. The full quantum information of this state
incorporates the information of all the individual function values
$f(i)$ but this is not accessible to any measurement. However
certain global properties of the collection of all the function
values {\em may} be determined by suitable measurements on
$\ket{f}$ which are not diagonal in the standard basis $\{
\ket{i}\ket{j} \}$. For example if $f$ is a periodic function, we may
determine the value of the period \cite{SH}, which falls far short
of characterising the individual function values but would
generally still require an exponential number of function
evaluations to obtain reliably by classical means.\\[3mm]
{\large\bf Entanglement Enhanced Information Processing}\\[3mm]
Suppose that we have a physical system of $n$ qubits in some
entangled state $\ket{\psi}$ and we apply a 1-qubit operation $U$
to the first qubit. This would count as one step in a quantum
computation (or rather a constant number of steps independent of
$n$, if $U$ needs to be fabricated from other basic operations
provided by the computer). Consider now the corresponding classical
computation. $\ket{\psi}$ may be described in components (relative
to the product basis of the $n$ qubits) by $a_{i_1 \cdots i_n}$
where each subscript is 0 or 1, and $U$ is represented by a
$2\times 2$ unitary matrix $U_i^j$. The application of $U$
corresponds to the matrix multiplication  \begin{equation}
\label{uu} a_{i_1
\cdots i_n}^{\rm (new)} = \sum_j U_{i_1}^j a_{j i_2 \cdots i_n}
\end{equation}
Thus the $2\times 2$ matrix multiplication needs to be performed
$2^{n-1}$ times, once for each possible value of the string $i_2
\cdots i_n$, requiring a computing effort which grows {\em
exponentially} with $n$. On a quantum computer, because of
entanglement, this $2^{n-1}$ repetition is unnecessary.

Consider now a unitary transformation $U$ of $n$ qubits (or more
precisely a family of such transformations labelled by $n$). $U$
may be described by a $2^n \times 2^n$ matrix and the computation
of $U\ket{\psi}$ classically by direct matrix multiplication
requires $O(2^n 2^n)$ operations. Even on a quantum computer $U$
needs to be fabricated (``programmed'') out of the basic operations
provided by the computer, each of which operate only on some
constant number of qubits. In general $U$ will require an
exponential number of such basic operations for its implementation.
It may be shown \cite{EJ,DD1} that $O(2^n 2^n)$ operations will
always suffice to program $U$ to any desired accuracy.

Suppose now that $U$ has the following special form. Let $c$ be any
constant, independent of $n$. Suppose that $U$ consists of the
sequential application of $p(n)$ unitary operations $V_i , i=1,
\ldots ,p(n)$ where each $V_i$ operates on only some $c$ out of the
$n$ qubits and $p(n)$ is a {\em polynomial} in $n$. An immediate
generalisation of the argument above shows that each $V_i$ may be
classically implemented (by matrix multiplication) in $O(c^2
2^{n-c} )= O(2^n )$ steps so that the classical computation of $U$
now requires $O(p(n)2^n )$ steps. This represents an exponential
saving over a general $U$ which required $O(2^n 2^n )$ steps but it
is still exponential in $n$. An important example of this partial
exponential speedup for classical computation is the so-called fast
Fourier transform algorithm \cite{MASROC}, as compared to the
regular Fourier transform algorithm. On a quantum computer each
$V_i$ requires some constant (independent of $n$) number of steps
to implement (programming the c-qubit operation $V_i$ in terms of
the basic operations) so that $U$ requires only $p(n)$ steps to
implement. In summary, if $U$ has the special form given above then
it still requires exponential time to compute classically (although
it {\em does} provide a partial exponential benefit here already)
but it requires only polynomial time to compute on a quantum
computer. Note however that after the quantum computation only a
small amount of information about the transformed data is
accessible to measurement, whereas the classical computation allows
the full information to be accessed.\\[3mm] {\large\bf The
Super-fast Quantum Fourier Transform}\\[3mm] The Fourier transform
on a finite Abelian group $G$ is a large unitary operation which
arises naturally in the mathematical formalism of group
representation theory. Furthermore it factorises in the special way
described in the previous section if the group has some additional
structure and it is known to be a basic tool for various useful
computational tasks, in particular the problem of determining
periodicity. Consequently, in view of the discussion above, it can
lead to quantum algorithms \cite{itpap,DJ,BV,SI,SH,EJ,KIT,GR} which
run substantially faster than any known classical algorithm for the
corresponding computational task. In this section we will outline
the construction of the Fourier transform and describe its
factorisation into unitary operations of a constant size.

Let $(G,+)$ be any finite Abelian group where we write the group
operation in additive notation. Let $|G|$ denote the number of
elements of $G$. An irreducible representation of $G$ is a function
\[ \chi : G\rightarrow C^* \] (where $C^*$ denotes the non-zero
complex numbers) satisfying
\begin{equation} \chi(g_1 +g_2 ) = \chi(g_1 ) \chi (g_2 )
\label{mult} \end{equation}
i.e. $\chi$ is a group homomorphism from the additive group $G$ to
the multiplicative group $C^*$. The condition eq. (\ref{mult}) has
the following consequences (see e.g. \cite{FUL,itpap} for proofs).
\begin{description}
\item[(A)] Any value $\chi (g)$ is a $|G|^{\rm th}$ root of unity.
Thus $\chi$ may be viewed as a group homomorphism $\chi
:G\rightarrow S^1$ where $S^1$ is the circle group of all unit
modulus complex numbers.
\item[(B)] Orthogonality (Schur's lemma): If $\chi_i$ and $\chi_j$
are any two such functions then:
\begin{equation} \label{ortho}
\frac{1}{|G|} \sum_{g \in G} \chi_i (g) \overline{\chi_j (g)} = \delta_{ij}
\end{equation}
(where the overline denotes complex conjugation).
\item[(C)] There are always exactly $|G|$ different functions $\chi$
satisfying eq. (\ref{mult}).
\end{description}
In view of (C) these functions may be exhaustively labelled by the
elements of $G$. Let $\{ \chi_g : g\in G \} $ be any such chosen
labelling. Then the Fourier transform on $G$ is the $|G|\times |G|$
matrix $\cal F$ whose rows are formed by listing the values of the
functions $\frac{1}{\sqrt{|G|}} \chi_g$:
\begin{equation} \label{FT}  {\cal F}_{gk} = \frac{1}{\sqrt{|G|}}
\chi_g (k) \hspace{1cm} g,k \in G \end{equation}
Note that by (B) $\cal F$ is always a {\em unitary} matrix.

In the context of quantum computation we will have a Hilbert space
$\cal H$ of dimension $|G|$ with a basis $\{ \ket{g}:g\in G \}$
labelled by the elements of $G$. Thus there is a natural shifting
action of $G$ on $\cal H$ given by
\begin{equation} \label{shift} U(k): \ket{g} \rightarrow \ket{g+k}
\hspace{1cm} k,g\in G \end{equation}
These operations all commute since $G$ is Abelian so there exists a
basis of simultaneous eigenstates of all the shifting operators.
According to (B) the states
\begin{equation} \label{states} \ket{\chi_k} = \frac{1}{\sqrt{|G|}}
\sum_{g\in G} \overline{\chi_k (g)} \ket{g} \hspace{1cm} k\in G
\end{equation}
form an orthonormal basis of $\cal H$ and using eq. (\ref{mult}) we
get \[ U(g) \ket{\chi_k} = e^{\chi_k(g)} \ket{\chi_k} \] so that
$\{ \ket{\chi_g} : g\in G\} $ is the basis of common eigenstates of
the shift operators. This basis is also called the Fourier basis.
The Fourier transform $\cal F$ is a unitary operation on $\cal H$
and using eq. (\ref{states}) with (\ref{FT}) and property (B) we
readily get:
\begin{equation} \label{bases} {\cal F} \ket{\chi_g} = \ket{g}
\end{equation}
so that the Fourier transform interchanges the standard and Fourier
bases.

Let $Z_q$ denote the additive group of integers mod $q$. It is well
known \cite{FRAL} that any finite Abelian group $G$ is isomorphic
to a direct product of the form
\begin{equation}\label{isom} G\cong Z_{m_1} \times Z_{m_2} \times
\ldots \times Z_{m_r} \end{equation}
(Furthermore we may require that $m_i$ divides $m_{i+1}$ and then
the numbers $m_i$ are unique). If we assume (usually without loss
of generality) that the group $G$ is presented as a product of the
form  eq. (\ref{isom}), then we can explicitly describe the
irreducible representations (\ref{mult}) and obtain a canonical
labelling of them by the elements of $G$. Suppose first that $G=
Z_m$. Consider the group homomorphism given by
\begin{equation} \label{tau}
\begin{array}{rcrl} \tau : & G\times G & \rightarrow & S^1 \\
 & (a,b) & \rightarrow & e^{2\pi i \frac{ab}{m}} \end{array}
 \end{equation}
It is easily verified that for each fixed $a\in G$ the function
$\chi_a :G\rightarrow S^1$ given by $\chi_a (b) = \tau (a,b)$
satisfies eq. (\ref{mult}) and there are $|G|$ such functions. Thus
we have obtained an explicit formula for the irreducible
representations, labelled in a natural way by the elements of $G$.
For the general case of a product $G= Z_{m_1} \times Z_{m_2} \times
\ldots \times Z_{m_r}$ we simply multiply the corresponding factors
in eq. (\ref{tau}) obtaining
\begin{equation} \label{taugen}
\begin{array}{rrrl} & \tau :  G\times G & \rightarrow & S^1 \\
 & ((a_1 , \ldots ,a_r ),( b_1 , \ldots , b_r))
 & \rightarrow & \exp 2\pi i \, (\frac{a_1 b_1}{m_1}+\frac{a_2 b_2}{m_2}
 \cdots +\frac{a_r b_r}{m_r})  \end{array}
 \end{equation}
and again \begin{equation} \label{chigen} \chi_{g_1}(g_2 ) = \tau(
g_1 ,g_2) \end{equation} provides the irreducible representations
labelled by the elements of $G$.

As an example consider the group $(Z_2)^n$ of all $n$-bit strings.
From eqs. (\ref{taugen}) (\ref{chigen}) and (\ref{FT}) we see that
the Fourier transform is just \[ {\cal F}_{\sigma \nu} =
\frac{1}{\sqrt{2^n}} e^{2\pi i \frac{\sigma \cdot \nu}{2}} =
\frac{1}{\sqrt{2^n}} (-1)^{\sigma \cdot \nu} \]
where $\sigma \cdot \nu = s_1 t_1 + \ldots + s_n t_n \mbox{ mod 2}$
if $\sigma = s_1 \ldots s_n$ and $\nu = t_1 \ldots t_n$. Thus in
this case the Fourier transform coincides with the Hadamard (Walsh)
transform. If $G= Z_{2^n}$ then we see using eqs. (\ref{tau}) and
(\ref{chigen}) that \[ {\cal F}_{ab} =
\frac{1}{\sqrt{2^n}} e^{2\pi i \frac{ab}{2^n}} \hspace{1cm} a,b= 0,
\ldots 2^n -1 \]
giving the familiar discrete Fourier transform modulo $2^n$.

As a unitary matrix the Fourier transform will act on vectors of
length $|G|$. We may view any such vector as a function
$f:G\rightarrow C$ on $G$ whose list of values $f(g_1 ), \ldots ,
f(g_{|G|}) $ defines the vector. The Fourier transform of $f$ is
then given by
\begin{equation} \tilde{f}(k) = \sum_{g\in G} {\cal F}_{kg} f(g)
= \frac{1}{\sqrt{|G|}} \sum_{g\in G} \chi_k (g) f(g)
\hspace{1cm} k\in G \label{FTGH} \end{equation}
We now describe the basic factorisation property of this large
unitary transformation which is necessary for its {\em efficient}
(i.e. polynomial time) implementation in the context of quantum
computation. The factorisation will be carried out relative to a
subgroup $H$ of $G$ and again the key ingredient will be the
property given by eq. (\ref{mult}). The basic technique was
developed by Cooley and Tukey \cite{COOTUK} leading to the
so-called fast Fourier transform (FFT) algorithm in classical
computation (which provides the partial exponential speedup noted
in the previous section) but the essential idea occurs already in
the work of Gauss \cite{GAUSS}.

Let $H$ be a subgroup of $G$ with index $I= |G|/|H|$. Let $k_1 +H,
k_2 +H, \ldots , k_I + H$ be a complete list of the cosets of $H$,
where $k+H \subseteq G$ denotes the subset given by $\{ k+h: h\in H
\} $. Thus $G$ is partitioned as a disjoint union $ (k_1 +H) \cup
(k_2 +H) \cup \ldots \cup (k_I +H) $. Hence the elements $g \in G$
may be written in a unique way in terms of the cosets as $g=k_i
+h$. Using eqs. (\ref{FTGH}) and (\ref{mult}) we get:
\begin{eqnarray} \tilde{f}(l) & = & \frac{1}{\sqrt{|G|}}\sum_{g\in G}
f(g) \chi_l (g) =
 \frac{1}{\sqrt{|G|}}\sum_{i=1}^I \sum_{h\in H} f(k_i +h) \chi_l
 (k_i + h) \nonumber \\
  & = & \frac{1}{\sqrt{|G|}}\sum_{i=1}^I \chi_l (k_i) \sum_{h\in H}
  f_i (h) \chi_l (h)
  \label{factor} \end{eqnarray}
where $f_i$ for $i=1, \ldots ,I$ are the functions on $H$ defined
by the restrictions of $f$ to the cosets: $f_i (h) = f(k_i +h)$.
The functions $\chi_l$ restricted to the subgroup $H$ satisfy eq.
(\ref{mult}) on $H$ so they are irreducible representations of $H$.
Hence the sum over $H$ in eq. (\ref{factor}) amounts to evaluating
the Fourier transform on $H$ of the functions $f_i$. Thus eq.
(\ref{factor}) expresses a decomposition of the Fourier transform
on $G$ into the evaluation of $I$ Fourier transforms on $H$ whose
results are then combined linearly in sums of length $I$ with
coefficients $\chi_l (k_i )$, done for each $l\in G$. Hence the
number of operations required is
\begin{equation} \label{tht} O( |H|^2 \times I + |G| \times I)
= O(|G|(|H|+I)) \end{equation}
where we have used $I=|G|/|H|$. This is generally better than the
$O(|G|.|G|)$ operations for the direct (matrix multiplication)
calculation of the Fourier transform on $G$. For example we choose
$H$ so that $I$ is small, say $I=2$ giving $|H|=|G|/2$ and then eq.
(\ref{tht}) represents an approximate halving of running time.

To enhance this benefit we iterate the construction on a {\em
tower} of subgroups \[ G\supset H_1 \supset H_2 \supset \cdots
\supset H_n \supset \{ 0 \} \] of greatest possible length,
ultimately expressing the Fourier transform of $G$ in terms of that
on the (small) subgroup $H_n$. An extensive survey of this
technique is given in \cite{MASROC}. We will illustrate it here
only for the group $Z_{2^n}$ and discuss the effect of the
resulting decomposition on the quantum computational
implementation. $Z_{2^n}$ has an optimal tower of subgroups with
each successive inclusion having the minimal possible index of 2:
\[ Z_{2^n} \supset Z_{2^{n-1}} \supset Z_{2^{n-2}} \supset \cdots
\supset Z_2 \supset \{ 0 \} \]
(Here $Z_{2^{n-1}}$ is the subgroup $\{ 0,2,4,\ldots ,2^n -2 \}$ of
all even integers in $Z_{2^n}$, $Z_{2^{n-2}}$ is the subgroup $\{
0,4,8,\ldots  \}$ of all multiples of 4 etc. and $Z_2$ is the
subgroup $\{ 0, 2^{n-1} \}$ ). Consider a general position $Z_{2^m}
\supset Z_{2^{m-1}}$ in this chain and let $FT_{2^m }$ denote the
Fourier transform on $Z_{2^m}$. The irreducible representations of
$Z_{2^m}$ are
\begin{equation} \label{charz}
 \chi_j (k) = (w^j )^k \hspace{1cm} \mbox{ for $j,k =0, \ldots 2^m -1$}
 \end{equation} where $w= \exp \frac{2\pi
i}{2^m}$. Then eq. (\ref{factor}) becomes (writing out the $i$-sum
explicitly):
\begin{eqnarray} \tilde{f}(j) & = & \sum_{k=0}^{2^m -1} f(k)
\frac{\chi_j (k)}{\sqrt{2^m}}
\nonumber \\  & = & \frac{1}{\sqrt{2}}\left( \sum_{k=0}^{2^{m-1}-1}
f(2k) \frac{w^{2jk}}{\sqrt{2^{m-1}}} + w^j \,
\sum_{k=0}^{2^{m-1}-1} f(2k+1) \frac{w^{2jk}}{\sqrt{2^{m-1}}} \right)
\label{Z2M}
\end{eqnarray}
Here the $f(2k)$ in the first sum and $f(2k+1)$ in the second sum
give the function $f$ restricted respectively to the cosets of
$Z_{2^{m-1}}\subset Z_{2^m}$ (i.e. the even and odd positions in
$Z_{2^m}$). Note that the irreducible representations of
$Z_{2^{m-1}}$ are the functions given in eq. (\ref{charz}) with $w$
replaced by $w^2= \exp \frac{2\pi i}{2^{m-1}}$. Thus the two
$k$-sums on RHS of eq. (\ref{Z2M}) are just $FT_{2^{m-1}}$ of the
even and odd labelled values of $f$. As $j$ in eq. (\ref{Z2M}) runs
through the values 0 to $2^{m}-1$, we cycle twice through the
$2^{m-1}$ components of the $FT_{2^{m-1}}$'s (noting that
$(w^2)^{2^{m-1}} =1$). If we restrict $j$ to running through the
values 0 to $2^{m-1}-1$ then $\tilde{f}(j)$ and
$\tilde{f}(j+2^{m-1})$ are both obtained from the $j^{\rm th}$
components of the two $FT_{2^{m-1}}$ transforms on RHS of eq.
(\ref{Z2M}), combined respectively with coefficients
$\frac{1}{\sqrt{2}}(1,w^j )$ and $\frac{1}{\sqrt{2}}(1,
w^{j+2^{m-1}})= \frac{1}{\sqrt{2}}(1,-w^j )$. Thus eq. (\ref{Z2M})
may be described as
\begin{equation} \label{m} \left. \begin{array}{ccl}
\tilde{f}(j) &= & \frac{1}{\sqrt{2}}( \mbox{ $j^{\rm th}$ cpt. of
$FT(f_{\rm even}) +w^j \cdot j^{\rm th}$ cpt. of $FT(f_{\rm odd})$
}) \\
\tilde{f}(j +2^{m-1}) & = & \frac{1}{\sqrt{2}}( \mbox{ $j^{\rm th}$ cpt. of
$FT(f_{\rm even}) -w^j \cdot j^{\rm th}$ cpt. of $FT(f_{\rm odd})$
}) \end{array} \right\} \end{equation} where $f_{\rm even}$ and
$f_{\rm odd}$ refer respectively to the $2^{m-1}$ even and odd
labelled values of $f$ and $j$ ranges from 0 to $2^{m-1}-1$. Now if
$C(2^m )$ denotes the number of operations required to
(classically) compute $FT_{2^m }$ then eq. (\ref{Z2M}) shows that
\[ C(2^m )= 2C(2^{m-1}) +O(2^m )
\] where the $O(2^m )$ arises from the extra additions and
multiplications needed for the $2^m$ $j$-values in eq. (\ref{Z2M}),
to linearly combine the results of the two $FT_{2^{m-1} }$
operations. The solution of this recursion relation is \[ C(2^n ) =
O(n2^n ) \] giving the partial exponential speedup (compared to
$O(2^n 2^n )$) noted previously.

In the context of quantum computation the data values $f(j)$ for
$j=0,\ldots , 2^{m} -1$ reside in the amplitudes of an entangled
state $\ket{f}$ of $m$ qubits. Writing $j$ in binary as an $m$ bit
string we have
\[ \ket{f}= \sum_{j_0 , j_1 , \ldots ,j_{m-1} =0}^{1}
\ket{f( j_{m-1}\ldots j_1 j_0 )} \ket{j_{m-1}}\cdots \ket{j_1}\ket{j_0} \]
and the qubits are numbered $0,1, \ldots ,m-1$ from right to left.
 The two $FT_{2^{m-1} }$ operations in eq.
(\ref{Z2M}), which operate on even and odd numbered components
respectively, may then be implemented by a {\em single}
$FT_{2^{m-1} }$ operation on qubits $m-1,m-2, \ldots ,1$ since the
values 0 and 1 of the remaining rightmost index respectively
determine the even and odd labelled positions (c.f. the discussion
of eq.(\ref{uu})). The $j^{\rm th}$ component of
$FT_{2^{m-1}}(f_{\rm even})$ (respectively $FT_{2^{m-1}}(f_{\rm
odd})$) then resides as the amplitude in dimension $2j$
(respectively $2j+1$). Thus to perform the linear recombination of
the two $FT_{2^{m-1}}$'s eq. (\ref{m}) shows that we need to
\begin{description}
\item[(a)] perform the unitary operation
\[ \frac{1}{\sqrt{2}} \left( \begin{array}{cc} 1 & w^j \\
1 & -w^j \end{array} \right) \] on dimensions $(2j,2j+1)$ for each
$j=0,\ldots , 2^{m-1}-1$.
\item[(b)] Reorder the answers according to the permutation $(2j,2j+1)
\rightarrow (j,j+2^{m-1})$ for each $j=0,\ldots , 2^{m-1}-1$ to get
$\tilde{f}(j)$ as the amplitude in dimension $j$.
\end{description}
This would appear to involve exponentially many operations (for the
$2^{m-1}$ values of $j$) but using the entanglement effects
discussed at eq. (\ref{uu}), we can achieve the result with only
$O(m)$ operations as follows. Note first that
\[ \frac{1}{\sqrt{2}} \left( \begin{array}{cc} 1 & w^j \\ 1 & -w^j
\end{array} \right) = \frac{1}{\sqrt{2}} \left( \begin{array}{cc}
 1 & 1 \\ 1 & -1 \end{array} \right) \left( \begin{array}{cc}
 1 & 0 \\ 0 & w^j \end{array} \right) \equiv H \cdot B_j \]
The operation $B_j$ in dimensions $(2j,2j+1)$ leaves the even
dimension unchanged and applies an $w^j$ phase shift in the odd
dimension. This may be achieved for all $j$ values {\em
simultaneously} by applying a 2-qubit gate $C_p$ to qubits 1 and
$p$ for each $p=1, \ldots , m-1$. Here $C_p$ is the conditional
phase shift of $w^{2^{p-1}}$ applied to qubit $p$ only if both
qubits 0 and $p$ are 1. In the standard basis of qubits 0 and $p$
we have:
\[ C_p = \left( \begin{array}{cccc} 1 & 0 & 0 & 0 \\ 0 & 1 & 0 & 0 \\
0 & 0 & 1 & 0 \\ 0 & 0 & 0 & w^{2^{p-1}} \end{array} \right) \]
Using the entanglement effects described at eq. (\ref{uu}) we see
that the successive application of the $m-1$ operations $C_p$
builds up a phase of $w^j$ in dimension $2j+1$ for each $j$. The
requirement that the zeroth qubit have value 1 selects the odd
positions and the conditional phase shift in $C_p$ builds up the
value $w^j$ successively for each `1' in the binary expansion of
$j$. All (exponentially many) values of $j$ with `1' in the $p^{\rm
th}$ place are treated simultaneously. Finally the 1-qubit
operation $H$ is applied just {\em once} to qubit 0, which
simultaneously applies $H$ to all pairs $(2j,2j+1)$ given by all
possible values of the remaining indices for qubits 1 to $m-1$
(c.f. eq. (\ref{uu})).

To implement (b) i.e. the permutation of dimension labels given
by\begin{center}  even labels: $2j \,\, \rightarrow \,\, j$
\hspace{1cm} odd labels: $(2j+1)
\,\, \rightarrow (j+2^{m-1})$ \end{center}
we simply cyclically permute the qubit labels as $m$-bit strings:
\[ i_{m-1} \ldots  i_1 i_0 \longrightarrow
i_0 i_{m-1} \ldots  i_1 \] If the label was even (i.e. $i_0
=0$) then the value is halved and if it was odd ($i_0 =1$) then the
cycling of $i_0$ to the leading position adds $2^{m-1}$ and the
residual even part is halved. This cycling may be physically
achieved by $m-1$ state swaps, of qubits 0 and 1, then 1 and 2 etc.
up to qubits $m-2$ and $m-1$. Alternatively we may just reorder the
output wires as shown in figure 1 below.
%
\setlength{\unitlength}{1cm}

\begin{picture}(13,11)(0,0)
\put(1.3,1){\line(1,0){0.9}}
\put(1.5,2){$\vdots$}
\put(1,3.2){\line(1,0){1.2}}
\put(1,4){\line(1,0){1.2}}
\put(1,5){\line(1,0){7.4}}
\put(2.2,0.6){\framebox(1.6,3.8){$FT_{2^{m-1}}$}}
\put(3.8,1){\line(1,0){3.7}}
\put(7.5,0.6){\framebox(1,0.8){\small $ C_{m-1}$}}
\put(8.5,1){\line(1,0){1.5}}
\put(10,1){\line(1,1){1}}
\put(11,2){\line(1,0){0.7}}
\put(11.8,1.9){\small $m-2$}
\put(3.8,3.2){\line(1,0){2.1}}
\put(5.9,2.8){\framebox(1,0.8){\small $ C_2$}}
\put(6.9,3.2){\line(1,0){3.1}}
\put(3.8,4){\line(1,0){0.8}}
\put(4.6,3.6){\framebox(1,0.8){\small $C_1$}}
\put(8.4,4.6){\framebox(0.8,0.8){$H$}}
\put(9.2,5.0){\line(1,0){0.8}}
\put(10,5){\line(1,-4){1}}
\put(11,1){\line(1,0){0.7}}
\put(11.8,0.9){\small $m-1$}
\put(5,4.4){\line(0,1){0.6}}
\put(4.8564,4.9){$\times$}
\put(6.3,3.6){\line(0,1){1.4}}
\put(8,1.4){\line(0,1){3.6}}
\put(5.6,4){\line(1,0){4.4}}
\put(10,4){\line(1,1){1}}
\put(10,3.2){\line(5,4){1}}
\put(11,5){\line(1,0){1}}
\put(11,4){\line(1,0){1}}
\put(6.1564,4.9){$\times$}
\put(7.8564,4.9){$\times$}
\put(12.1,4.9){0}
\put(12.1,3.9){1}
\put(11.5,2.7){$\vdots$}
\put(7.2,2){$\vdots$}
\put(0.2,0.9){\small $m-1$}
\put(0.6,4.9){0}
\put(0.6,3.9){1}
\put(0.6,3.1){2}
\put(12.1,4.9){0}
\put(12.1,3.9){1}
\put(1,8.8){\line(1,0){1}} \put(0.6, 8.7){2}
\put(1,9.6){\line(1,0){1}} \put(0.6,9.5){1}
\put(1,10.4){\line(1,0){1}}  \put(0.6,10.3){0}
\put(2,7){\framebox(2,4){$FT_{2^m}$}}
\put(4,7.4){\line(1,0){2.7}} \put(6.8,7.3){\small $m-1$}
\put(4,8.8){\line(1,0){3}} \put(7.1,8.7){2}
\put(4,9.6){\line(1,0){3}}  \put(7.1,9.5){1}
\put(4,10.4){\line(1,0){3}}  \put(7.1,10.3){0}
\put(9,9){EQUALS:}
\put(1.5,7.8){$\vdots$} \put(5,7.8){$\vdots$}
\put(1.3,7.4){\line(1,0){0.7}} \put(0.2,7.3){\small  $m-1$}
\end{picture}
%

\noindent
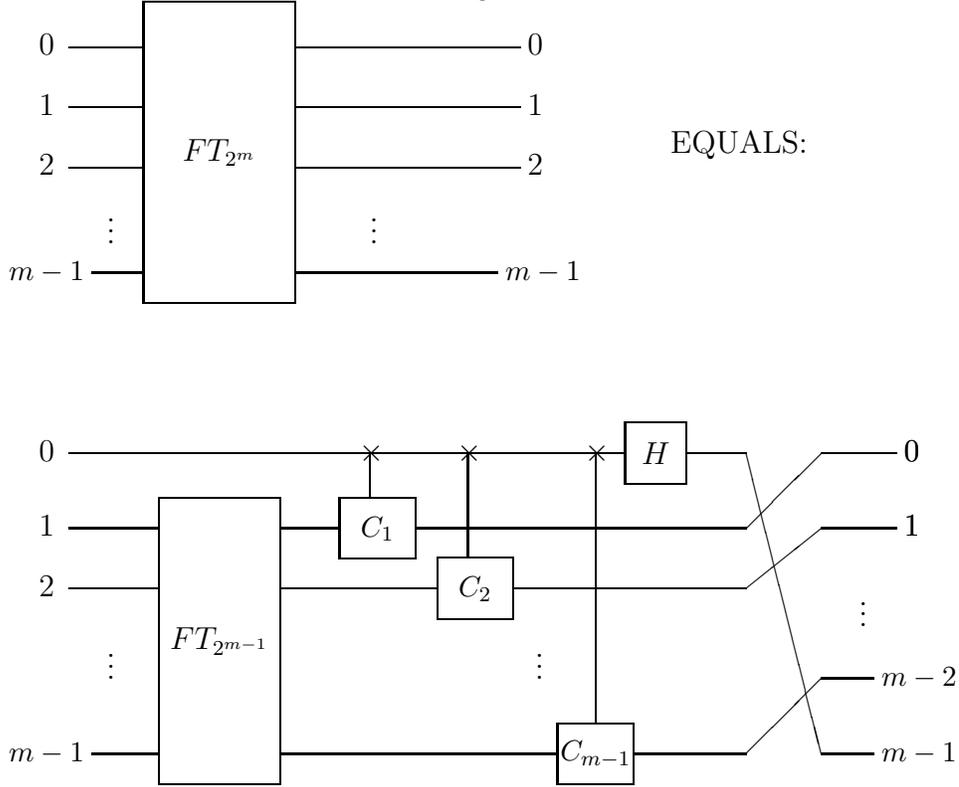
{\bf Figure 1.} The network diagram for the decomposition of
$FT_{2^m}$ into $FT_{2^{m-1}}$ and $O(m)$ extra operations. The
conditional phase shift $C_p$ on the $p^{\rm th}$ qubit is denoted
by a box on the $p^{\rm th}$ qubit line and a connection across to
the $0^{\rm th}$ qubit with a cross to denote the fact that its
operation on the $p^{\rm th}$ qubit is ``controlled'' by the
requirement that the $0^{\rm th}$ qubit have value 1.\\[3mm]

Iterating this construction for $FT_{2^{m-1}}$ in terms of
$FT_{2^{m-2}}$ etc. yields the standard network for the fast
Fourier transform on $Z_{2^n}$ as given for example in \cite{EJ}.

If $Q(2^m )$ denotes the number of operations needed to implement
$FT(2^m )$ in the quantum context then the above description shows
that
\[ Q(2^m )= Q(2^{m-1}) + O(m) \]
giving \[ Q(2^n ) = O(n^2 ) \] This quadratic time quantum
algorithm for $FT(2^n )$ is used in Shor's factoring algorithm
\cite{EJ,itpap}.
\\[3mm]
{\large\bf Utility of the Fourier Transform}
\\[3mm]
The utility of the Fourier transform $\cal F$ in the algorithms of
Deutsch, Simon and Shor \cite{DJ,SI,SH} has been described in
\cite{itpap}. We will here outline in general terms, its
fundamental application to the determination of periodicities. A
different interpretation of $\cal F$ in terms of the problem of
phase estimation, has been given in \cite{CEM}.

Let $f:G\rightarrow X$ be a function on the group (taking values in
some set $X$) and consider
\[ K= \{ k\in G : f(k+g) = f(g) \mbox{ for all $g\in G$} \} \]
$K$ is necessarily a subgroup of $G$ called the stabiliser or
symmetry group of $f$. It characterises the periodicity of $f$ with
respect to the group operation of $G$. Given a device that computes
$f$, our aim is to determine $K$. More precisely we wish to
determine $K$ in time $O({\rm poly}(\log |G|))$ where the
evaluation of $f$ on an input counts as one computational step.
(Note that we may easily determine $K$ in time $O({\rm poly}(
|G|))$ by simply evaluating and examining all the values of $f$).
We begin by constructing the state
\[ \ket{f}= \frac{1}{\sqrt{|G|}} \sum_{g\in G} \ket{g}\ket{f(g)}  \]
and read the second register. Assuming that $f$ is suitably non-degenerate
-- in the sense that $f(g_1 ) = f(g_2 )$ iff $g_1 - g_{2} \in K$ i.e.
that $f$ is one-to-one within each period -- we will obtain in the first
register
\begin{equation} \label{per}
\ket{\psi (g_0 )} = \frac{1}{\sqrt{|K|}} \sum_{k\in K} \ket{g_0 +k}
\end{equation}
corresponding to seeing $f(g_0 )$ in the second register and $g_0$
has been chosen at random. In eq. (\ref{per}) we have an equal
superposition of labels corresponding to a randomly chosen coset of
$K$ in $G$. Now $G$ is the disjoint union of all the cosets so that
if we read the label in eq. (\ref{per}) we will see a random
element of a random coset, i.e. a label chosen equiprobably from
all of $G$, yielding no information at all about $K$. The Fourier
transform will provide a way of eliminating $g_0$ from the labels
which may then provide direct information about $K$. Consider the
basis $\{
\ket{\chi_g}: g\in G
\} $ of shift invariant states introduced in eq. (\ref{states}).
 Next note that the state in eq.
(\ref{per}) may be written as a $g_0$-shifted state:
\[ \sum_{k\in K} \ket{g_0 + k} = U(g_0 )\left( \sum_{k\in K} \ket{k} \right) \]
Hence if we write this state in the basis $\{ \ket{\chi_g}, g\in G
\}$ then $\sum_k \ket{k}$ and $\sum_k \ket{g_0 +k}$ will contain the
same pattern of labels, determined by the subgroup $K$ only.
According to eq. (\ref{bases}) the Fourier transform converts the
shift-invariant basis into the standard basis.  Thus after applying
$\cal F$ to eq. (\ref{per}) we may read the shift-invariant basis
label by reading in the standard basis, yielding information about
$K$.

In terms of the presentation of $G$ given in eq. (\ref{isom})  and
the associated formulas for the irreducible representations given
by eqs. (\ref{taugen}) and (\ref{chigen}) we may compute explicitly
the pattern of labels associated with a subgroup $K\subset G$. As
an example consider $G=Z_{mn}$ and $K=mZ=\{ 0,m,2m,\ldots ,(n-1)m
\} $ with $|K|= n$. Then the Fourier transform of the fundamental
periodic state $\ket{K} = \frac{1}{\sqrt{n}} \sum_{k\in K} \ket{k}$
is \begin{equation} \label{star} {\cal F}\ket{K} =
\frac{1}{n\sqrt{m}}
\sum_{l\in G}
\left( \sum_{k\in K} \chi_l (k) \right) \ket{l} \end{equation}
Thus the labels appearing are precisely those $l\in Z_{mn}$ for
which \begin{equation} \label{neq} \sum_{k\in K} \chi_l (k) \neq 0
\end{equation} To sort out this condition we introduce a further
elementary property of irreducible representations. For any group
$G$ the constant function $\chi (g)=1$ for all $g\in G$, is clearly
an irreducible representation (the trivial representation) and
using the orthogonality property (B) between $\chi$ and any {\em
other} irreducible representation $\chi'$ we see that
\[ \sum_{g\in G} \chi' (g) =0 \]
Now $\chi_l$ restricted to the subgroup $K$ is an irreducible
representation of $K$ so eq. (\ref{neq}) can hold if and only if
\[ \chi_l (k) =1 \hspace{1cm} \mbox{ for all $k\in K$} \]
According to eqs. (\ref{taugen}) and (\ref{chigen}) we have
\[ \chi_l (k) = \exp 2\pi i \frac{kl}{mn} = \exp 2\pi i \frac{cl}{n} \]
where we have introduced $c$ using the fact that $k=cm$ is always a
multiple of $m$, by definition of $K$. This will equal 1 for all
$c=0,
\ldots , (m-1)$ if and only if $l$ is a multiple of $n$ i.e. $l=
0,n,2n,
\ldots ,(m-1)n$. Thus the pattern of labels associated with $mZ
\subset Z_{mn}$ is $nZ$ and furthermore in eq. (\ref{star}) each
such label will appear with equal amplitude $\frac{1}{\sqrt{m}}$. A
similar calculation for the subgroup $\{ 0,\xi
\} \subset (Z_2 )^n$ (where $\xi$ is a chosen $n$-bit string) shows that
the resulting pattern of labels, after applying the Fourier
transform for $(Z_2 )^n$ to the periodic state $\frac{1}{\sqrt{2}}(
\ket{0}+\ket{\xi})$, is $\{ \nu :
\xi \cdot \nu = 0 \} $. This fact forms the basis of Simon's algorithm
\cite{SI,itpap,BH}.
\\[3mm]
{\large\bf Conclusion}
\\[3mm]
Let $\ket{\psi}$ be an $n$-qubit entangled state and $U$ a 1-qubit
unitary operation.We have seen that the one-step physical operation
of applying $U$ to (say) the first qubit of $\ket{\psi}$
corresponds to a state transformation which generally requires an
exponential (in $n$) effort to compute classically. Indeed
mathematically the transformation is represented by a {\em tensor}
product $U\otimes I_2 \otimes \ldots \otimes I_2 $ (where $I_2$ is
the 2 by 2 identity matrix, which represents the operation of
``doing nothing'' on the corresponding qubits). The tensor product
spreads the effect of $U$ into an exponentially large matrix.
Stated otherwise, we can say that the physical operation of {\em
doing nothing} to a subsystem of an entangled system is a highly
nontrivial operation and gives rise to an exponentially enhanced
information processing capability (when performed in conjunction
with some operation on another small part of the system).

We have given an analysis of the implementation of the fast Fourier
transform algorithm in a quantum context and shown that its
exponential speedup (as compared to the corresponding classical
computation) derives wholly from the above tensor product property.
We have also given a general discussion of the role of entanglement
in quantum computation and the utility of the Fourier transform in
the known quantum algorithms.
\\[3mm]
{\large\bf Acknowledgements}
\\[3mm]
This work was supported in part by the European TMR Research
Network ERB-FMRX-CT96-0087 and the National Institute for
Theoretical Physics at the University of Adelaide.

\end{document}